\documentclass{article}    

\usepackage{graphicx}
\usepackage{amsfonts,amssymb,amsbsy,amsmath,mathrsfs,enumerate,verbatim,amsthm}
\usepackage{mathptmx}
\usepackage{helvet}
\usepackage{courier}        
\usepackage{type1cm}
\usepackage{makeidx}         
\usepackage{multicol}
\usepackage[bottom]{footmisc}
\usepackage{epsf}
\usepackage{epstopdf}
\usepackage{pdfsync}
\usepackage{epsfig}
\usepackage{algorithmic,algorithm}
\usepackage{hyperref}
\usepackage{color}  
\usepackage{afterpage}
\usepackage{verbatim}

\makeindex

\allowdisplaybreaks

\begin{document}

\title{ Frequency inversion   method and device for malignant melanoma detection using RF/microwaves}

\author{L.~Beilina  \thanks{Department of Mathematical Sciences, Chalmers University of Technology and University of Gothenburg, SE-42196 Gothenburg, Sweden, e-mail:
\texttt{\
larisa@chalmers.se}}
\and
A.~Eriksson  \thanks{Department of Mathematical Sciences, Chalmers University of Technology and University of Gothenburg, SE-42196 Gothenburg, Sweden, e-mail:
\texttt{\
 andfh@chalmers.se}}
\and
N. Neittaanmäki  \thanks{Department of Laboratory Medicine, Institute of Biomedicine, University of Gothenburg , e-mail:
\texttt{\
noora.neittaanmaki@gu.se; the paper is published in the conference proceedings of   2024 International Conference on Electromagnetics in Advanced Applications (ICEAA), Lisboa, Portugal, IEEE}}
 }

\date{\today}

\maketitle

\graphicspath{
  {FIGURES/}
 {pics/}}

\begin{abstract}
 The paper considers an inverse problem of reconstructing the
  spatially distributed complex dielectric permittivity function
  using backscattered data of the electric field in the frequency domain.
  We develop iterative algorithm for reconstruction of this function.
 Numerical example  show the performance of
 the proposed algorithm in 2D.
 Additionally, we  present
 at the first time
 the  new prototype of the device for the 
 malignant melanoma
 detection
 which is modelled in HFSS/Ansys.
 Results of measurements of scattered parameters
 in HFSS/Ansys show
significant phase distortion in the case when melanoma is present compared to the case
 when melanoma is absent.
\end{abstract}

\vspace{0.5cm}


\vspace{0.5cm}


\section{Introduction}

In this work is developed reconstruction algorithm for the problem of
determination of the spatially distributed complex dielectric
permittivity function using scattered frequency-dependent data of the
electric field collected in  the investigated domain.
 Additionally, we  present results of the ongoing work
 on data acquisition procedure using prototype of the
 device modelled in HFSS/Ansys for malignant melanoma (MM) detection
 using RF/microwaves.
 Malignant melanoma is the
 deadliest of skin cancers. The prognosis of MM is related to the
 primary tumors invasion depth in the skin \cite{skindepth}.
 Microwave medical imaging
 is non-invasive imaging \cite{Pastorino}.
 Thus, it is very attractive
 addition to the existing medical imaging technologies like  ultrasound \cite{gonch2}, X-ray  and MRI imaging \cite{tomography}. However,  X-ray and MRI are not used
 in diagnosis of primary skin cancers, and other 
 non-invasive methods which can be used directly on the skin like OCT, RCM,
 are applied  \cite{noninvasive}.

In  \cite{ieee1,mri} were reported different
 malign-to-normal tissues contrasts, revealing that malign
 tumors have  higher relative
 permittivity  values, than normal tissues.
 The challenge is to accurately estimate the relative
permittivity of the internal structures using the information from the
backscattered electromagnetic waves of frequencies around 1 GHz
collected at several detectors.

 Since the 90-s quantitative reconstruction algorithms for medical
 imaging based on the solution of Maxwell's system have been developed
 to provide images of the complex permittivity function, see
 \cite{ieee10} for 2D techniques, \cite{hyperthermi,BAK,EBL,ieee15,ieee12,litman1} for 3D
 techniques in the frequency domain and \cite{ieee19, ieee,BL1,BL2,BondestaB}
 for time domain (TD) techniques.  In all above cited works microwave
 medical imaging remained the research field and had little clinical
 acceptance \cite{meaney1}.

Potential application of the algorithm  and device developed in this work
is  in  malignant melanoma detection using microwaves.
In numerical example of this work we
will focus on microwave medical imaging of realistic
skin phantom taken from online ISIC repository \cite{isic}.
In this work we
 tested
 simplified version of the
 proposed reconstruction  algorithm which allows
 determine the dielectric permittivity function under the condition
 that the effective conductivity function is known.  Currently we are
 working on  determination of both
 spatially distributed functions, dielectric permittivity and conductivity.

 The numerical test of the current work shows that
the  proposed iterative algorithm  can efficiently and accurately
reconstruct the relative dielectric permittivity function
using measured data of the electric field at some frequency interval.
Additionally,  we present at the first time
the  new prototype of the
device modelled in HFSS/Ansys    for MM detection.
Results of measurements of scattered parameters using developed
   prototype show
significant phase distortion in the case when melanoma is present compared to the case
 when melanoma is absent.

\section{Statements of forward and inverse problems}
\label{sec:model}

Let $ \Omega \subset \mathrm{R}^d, d=2,3$ be a convex
domain   supplied by
Silver-Muller radiation condition at infinity.
 Our basic model equation is
  the following vector wave equation  for electric field $\widehat{E}\left( x, \omega\right) =\left(
\widehat{E}_{1},\widehat{E}_{2}, \widehat{E}_{3}\right) \left( x,
\omega \right)$:
 \begin{equation}\label{rotrot1}
 \nabla \times \nabla \times \widehat{E}(x,\omega) -
 \omega^2  \varepsilon'(x) \widehat{E}(x,\omega) =  -i  \omega \mu_0 \widehat{J}(x,\omega).
 \end{equation}
 Here, $\varepsilon'(x)$ is the spatially distributed complex dielectric function
 defined as
\begin{equation}\label{rotrot2}
    \varepsilon'(x) =
    \varepsilon_r(x)\frac{1}{c^2} + i \mu_0 \frac{\sigma(x)}{\omega},
\end{equation}
where $\omega$ is the angular frequency,
 $\varepsilon_r(x) = \varepsilon(x)/\varepsilon_0$
and $\sigma(x)$ are the dimensionless relative dielectric
permittivity and electric conductivity functions (S/m), respectively,
$\varepsilon_0, \mu_0$ are the permittivity and permeability of the
free space, respectively,
 $c=1/\sqrt{\varepsilon_0 \mu_0}$ is the speed of
light in free space,  $\widehat{J}(x,\omega)$ is the known source function.


Applying transformation  $ \nabla \times \nabla \times
\widehat{E} = \nabla ( \nabla \cdot \widehat{E}) - \nabla \cdot
(\nabla \widehat{E})$ in \eqref{rotrot1}
and assuming $\varepsilon = \textrm{const}$ in $\Omega$ 
we obtain inhomogeneous vector Helmholtz
equation
\begin{equation}\label{helmholtz}
\begin{split}
\triangle  \widehat{E} + k^2 \widehat{E} &= i \omega \mu_0  \widehat{J},
\end{split}
\end{equation}
 where $k^2 = \omega^2  \varepsilon'$.

 To formulate an inverse problem we
 consider now bounded domain $\Omega$  with a boundary $\partial \Omega$
 and
 assume that for some known constants $M_1 > 1, M_2 > 0$, the functions
$\varepsilon_r(x)$ and $\sigma(x)$ of equation \eqref{rotrot2} satisfy
conditions
\begin{equation} \label{2.3}
\begin{split}
  \varepsilon_r(x) &\in \left[ 1, M_1\right],\quad
  \sigma(x) \in \left[0, M_2\right], ~\text{ for } x\in  \Omega, \\
  ~  ~ \varepsilon_r(x) &=1, \quad \sigma(x) = 0, ~~ \mathbb{R}^{3}\backslash \Omega.
\end{split}
\end{equation}

According to the Tikhonov's theory \cite{tikhonov} we always know a
priori the constants $M_1,M_2$. In other words, we know the set of
admissible parameters for $\varepsilon_r(x)$ and $\sigma(x)$.

Let us formulate the following inverse problems:

 \textbf{Inverse  Problem IP1 (boundary measurements)} \emph{ Assume that functions $\varepsilon_r(x), \sigma(x)$ of the real and imaginary parts of the function
}$\varepsilon'\left(x\right)$ \emph{\  in \eqref{rotrot2}    are  unknown in
  the domain }$\Omega$\emph{. Determine
  these functions} \emph{\ for }$x\in \Omega,$ \emph{\ assuming that the following
  function }$\tilde{E}(x,\omega) $\emph{\ is known  for} $\omega=\bar{\omega}$
\begin{equation}\label{ip1}
  \widehat{E}(x, \bar{\omega}) = \tilde{E}(x, \bar{\omega}) , \forall x
  \in  \partial \Omega.
\end{equation}

\textbf{Inverse  Problem IP2 (internal measurements)} \emph{ Assume that functions $\varepsilon_r(x), \sigma(x)$ of the real and imaginary parts of the function
}$\varepsilon'\left(x\right)$ \emph{\  in \eqref{rotrot2} are  unknown in
  the domain }$\Omega$\emph{. Determine
  these functions}  \emph{\ for }$x\in \Omega,$ \emph{\ assuming that the following
  function }$\tilde{E}(x,\omega) $\emph{\ is known  for} $\omega=\bar{\omega}$
\begin{equation}\label{ip2}
  \widehat{E}(x, \bar{\omega}) = \tilde{G}(x, \bar{\omega}) , \forall x
  \in  \Omega.
\end{equation}

The functions $\tilde {E}\left(x,\omega\right) $ in (\ref{ip1}) and $\tilde{G}(x, \bar{\omega})$  in
   \eqref{ip2}  represent
the frequency-dependent measurements of all
components of the electric wave field
$\widehat{E}(x,\omega)= (\widehat{E_1}, \widehat{E_2},\widehat{E_3})(x,\omega)$
  for $\omega=\bar{\omega}$.

Further, let us denote the standard inner product in $[L^2(\Omega)]^d$ as
$(\cdot,\cdot), ~ d \in \{1,2,3\}$, and the corresponding norm by
$\parallel \cdot \parallel$. The inner  product in space and time is denoted by
$((\cdot,\cdot))_{\Omega_T}, ~ d \in \{1,2,3\}$.


\subsection{Finite element method for  reconstruction of complex dielectric permittivity function}
\label{sec:fem}

In this section we explain how  to solve \textbf{IP2}  and reconstruct the function $\varepsilon'(x)$
of the equation (\ref{helmholtz}) using the variational formulation of the following model
equation
 considered in the computational domain $\Omega$:
\begin{equation}  \label{2.7n}
  \begin{split}
\Delta  \widehat{E}(x,\omega) + \omega^{2} \varepsilon'(x)  \widehat{E}(x, \omega)  &= 0,\text{ }x\in
\Omega,  \\
\partial_n \widehat{E}(x,\omega)  &=  \psi(x,\omega),   \text{ }x\in  \partial \Omega.
\end{split}
\end{equation}
Here, the function  $\psi(x,\omega)$  is
 known  at the boundary $\partial \Omega$. It can be approximated by knowing $\tilde{E}$ in the neighborhood of the boundary of the computational domain.
The equation  \eqref{2.7n}
is obtained from  \eqref{helmholtz}  under the condition that
 the source function is outside of the computational domain $\Omega$, or
$  \widehat{J}  \notin \overline{\Omega }$.

 In order to develop iterative algorithm we choose
 frequency interval
  $[ \underline{\omega}, \bar{\omega}]$ 
 and make a partition of this interval into $N$ sub-intervals
$\bar{\omega}= \omega_{N} > \omega_{N-1}> \cdots  > \omega_{0}=\underline{\omega}$ such that
\begin{equation}\label{discrete}
  \underline{\omega}= \omega_{0} < \omega_{1}<...< \omega_{N-1} < \omega_{N}=\overline{\omega},\omega_{n} -
   \omega_{n-1}= \delta \omega,
\end{equation}
where $\delta \omega$ is the step size of every interval  and $\varepsilon'_n(x) = \varepsilon'(x)$
for $ \omega \in ( \omega_n, \omega_{n+1}]$.
  Thus, we  will solve now the equation \eqref{2.7n} on every frequency interval
$[ \omega_{n}, \omega_{n+1}]$.

Let us now compute
$\varepsilon'_{n}$ on every   frequency interval
$[ \omega_{n}, \omega_{n+1}]$ using  the finite
element   formulation of the problem (\ref{2.7n}).
To do this we 
discretize  the  domain
$\Omega$ into elements $K$. Next,
 we  denote by $K_h = \{K\}$ a set such that
\begin{equation} \label{eq:fe.mesh}
 K_h = \cup_{K \in K_h} K=K_1 \cup K_2...\cup K_l,
\end{equation}
 where $l$ is the
 number of elements $K$ in $\overline{\Omega}$.
 Here, $h$ is the  mesh function
 defined as 
\begin{equation}\label{meshfunction}
h |_K = h_K ~~~ \forall K \in K_h
\end{equation}
and denoting the local diameter of $K$ which we define as the longest side of $K$.
We make also assumption on minimal angle condition for the elements $K$ in the finite element mesh $K_h$ \cite{KN}.

To approximate the functions
 $\widehat{E}_n(x) = \widehat{E}(x) $
for $ \omega \in ( \omega_n, \omega_{n+1}]$
we introduce the
  finite element  space $V_h$ defined by
\begin{equation}\label{eq:fe.p1}
V_h := \{ u \in  H^1(\Omega): u|_{K} \in  P_1(K),  \forall K \in K_h \}.
\end{equation}
Here, $P_1(K)$ denotes the set of piecewise linear functions on each element $K$
 in the finite element mesh $K_h$.
To approximate $\varepsilon'_n(x)$ ,
   we define the space of piecewise constant functions $C_{h} \subset L_2(\Omega)$, 
\begin{equation}\label{p0}
C_{h}:=\{u\in L_{2}(\Omega ):u|_{K}\in P_{0}(K),\forall K\in  K_h\}, 
\end{equation}
where $P_{0}(K)$ are the piecewise constant functions on $K$.
Then, the finite element formulation for (\ref{2.7n})  reads:
Find $\varepsilon'_n \in C_h$  such that   for the known functions    $ \widehat{E}_n,\psi_n$ and for all $v \in V_h$, $ \omega = \omega_n$ on every
 frequency interval
$ \left( \omega_{n-1},\omega_{n}\right]$, we have
\begin{equation}
  ( \varepsilon'_n  \widehat{E}_n, v) =  \frac{1}{\omega_{n}^{2}}(\nabla \widehat{E}_n, \nabla  v )
   - \frac{1}{\omega_{n}^{2}} (\psi_n, v )_{\partial \Omega}.
  \label{3.107_1}
\end{equation}

We expand $\widehat{E}_n  $ in   \eqref{3.107_1} using the standard continuous piecewise
linear functions $\{\varphi_k\}_{k=1}^N$ in space as 
\begin{equation}\label{new}
 \widehat{E}_n =\sum_{k=1}^N \widehat{E}_{n,k} \varphi_k(x),
\end{equation}
 where $\widehat{E}_{n,k} $ are the discrete nodal values of the
 already computed functions  $ \widehat{E}_n$, or  measured data $\tilde{G}_n$,
 for every frequency $ \omega = \omega_n$ on the
 frequency interval
$ \left( \omega_{n-1},\omega_{n}\right]$.
We substitute the 
expansion (\ref{new}) in the variational formulation (\ref{2.7n})
with $v(x) = \varphi_j(x)$, and obtain the following system of discrete equations
\begin{equation}
  \begin{split}
\sum_{k,j=1}^N ( \widehat{E}_{n,k}~ \varphi_k, \varphi_j)
\varepsilon'_{n,k} &=  \frac{1}{\omega_{n}^{2}} \sum_{k,j=1}^N( \nabla \varphi_k, \nabla  \varphi_j )  \widehat{E}_{n,k} - \\
&\frac{1}{\omega_{n}^{2}} \sum_{k,j=1}^N  (   \psi_{n,k} , \varphi_j)_{\partial \Omega}.  \label{3.108}
\end{split}
\end{equation}
We can write the system (\ref{3.108})  in the matrix form for the unknown
$\varepsilon'_n $ and known $\widehat{E}_n, \psi_n$
 for every frequency $ \omega = \omega_n$ on the
 frequency interval
$ \left( \omega_{n-1},\omega_{n}\right]$
as
\begin{equation}
M  \varepsilon'_n  =  \frac{1}{\omega_{n}^{2}}  G  \widehat{E}_{n}  -  \frac{1}{\omega_{n}^{2}} F_n.
\label{3.108_1}
\end{equation}
Here, $\widehat{E}_n$ is the vector of known data obtained on the interval $[ \omega_{n-1}, \omega_{n}]$,
$M$ is the block mass matrix in space  with local matrix entries
\begin{equation*}
  M_{n}^{K}  =    (\widehat{E}_{n,k} ~\varphi_k, \varphi_j)_K, 
 \end{equation*}
$G$ is the stiffness
matrix corresponding to the gradient term in  (\ref{3.108}) 
with local matrix entries
 \begin{equation*}
  G_{n}^{K}  =   (\nabla \varphi_k, \nabla  \varphi_j)_K,
 \end{equation*}
and $F_n$ corresponds to the load vector with local vector entries
 \begin{equation*}
  F_{n}^{K} = ( \psi_{n,k}, \varphi_j)_{\partial K \in \partial \Omega}.
 \end{equation*}

To
obtain an explicit scheme for the computation of the complex dielectric permittivity
function
 $\varepsilon'_n $  we approximate $M$ by the lumped mass matrix $M^{L}$ in
space and get the following equation for the
explicit computation of the function $\varepsilon'_n$ in (\ref{2.7n})
on every   frequency interval
$[ \omega_{n-1}, \omega_{n}]$  :
\begin{equation}
\varepsilon'_n  = \frac{1}{\omega_{n}^{2}}  (M^{L})^{-1}  G \widehat{E}_{n}   - \frac{1}{\omega_{n}^{2}}  (M^{L})^{-1}  F_n.
  \label{3.109}
\end{equation}

Algorithm 1 summarizes iterative procedure for computation of the
function $\varepsilon'(x)$.
We note that  approximation of $M$ by the lumped mass matrix $M^{L}$
is very important feature of this algorithm.

\vspace{0.2cm}

\begin{algorithm}[hbt!]
  \centering
  \caption{ Variational  frequency inversion   algorithm  }
    \begin{algorithmic}[1]


      \STATE Initialization: choose the finite element mesh $K_h$ in $\Omega$.
     Choose and discretize the frequency interval $[ \underline{\omega},
  \bar{\omega}]$ into $N$ sub-intervals as in \eqref{discrete}.


   \STATE  For  iterations $n=1,\,2,\,\ldots ,\,N$

\begin{itemize}
 \item   Compute $\widehat{E}_{n}$ or obtain measured data $\tilde{G}_n$ on $K_h$   for $\omega= \omega_n$. Compute $\psi_n$.

 \item  Compute  approximations  of $\varepsilon'_{n}$ via  \eqref{3.109}.

 \end{itemize}

     \STATE Go to the next frequency interval $\left[ \omega_{n},\omega_{n+1}\right] $ if $n<N$
     and repeat all steps in item 2.

   \STATE  Stop computing approximations $\varepsilon'_n$   of the functions ${\varepsilon}'_{n}$    when either $e_n \geq e_{n-1}$ or $e_n \leq
tol$, where the relative error $e_n$
 is computed as
\begin{equation}\label{An}
  e_n: =
  \frac{ \|~ |\varepsilon'_n| - |\varepsilon'_{n-1}|~ \|_{L_2(\Omega)}}{\|~ |\varepsilon'_n| ~\|_{L_2(\Omega)}}
  \end{equation}   
for a chosen tolerance $tol$, or when  $e_n$ starts abruptly  grow. Here,
  $|\varepsilon'_n| = \sqrt{Re( \varepsilon'_n)^2 + Im(\varepsilon'_n)^2}$.

  \STATE  If $n=N$, then stop.

 \end{algorithmic}
\end{algorithm}

\begin{figure}[tbp]
 \begin{center}
   \begin{tabular}{c}
     \includegraphics[trim = 0.0cm 0.0cm 0.0cm 0.0cm, scale=0.7, clip=]{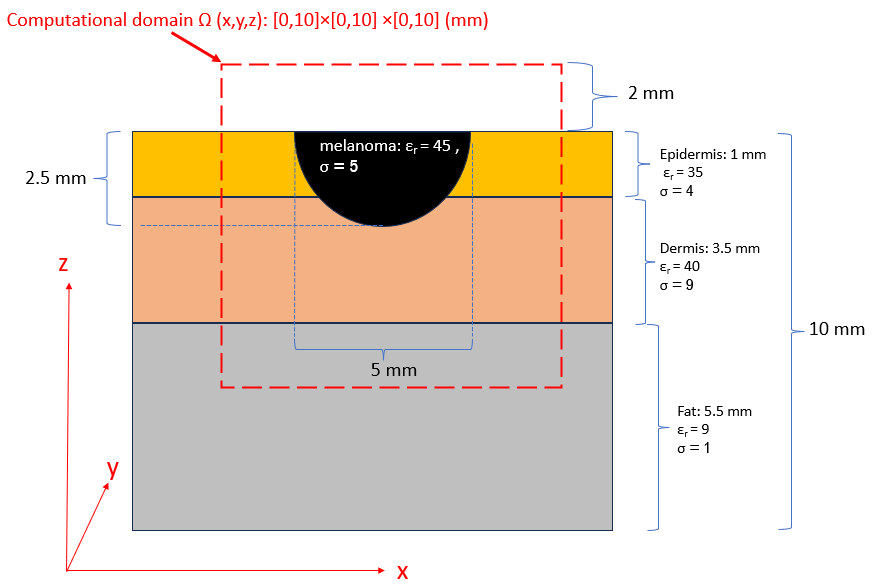}
\end{tabular}
\end{center}
 \caption{\small
 Computational domain in data  acquisition in 3D. }
\label{fig:model3D}
\end{figure}

\begin{figure}[tbp]
 \begin{center}
   \begin{tabular}{cc}
     \includegraphics[trim = 0.0cm 0.0cm 0.0cm 0.0cm,   scale=0.4,clip=]{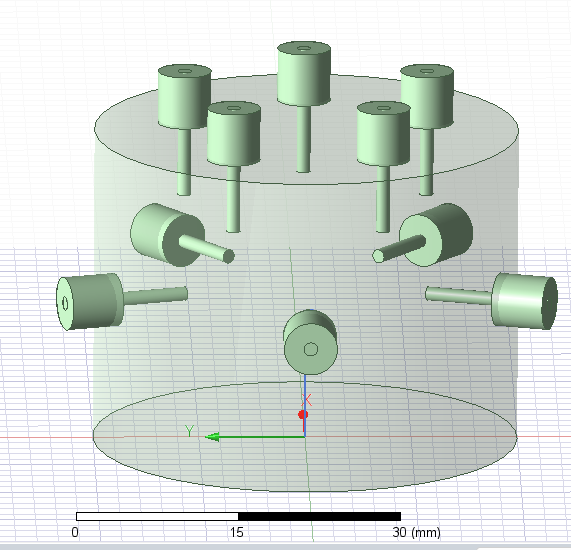} &
      \includegraphics[trim = 0.0cm 0.0cm 0.0cm 0.0cm,   scale=1.2,clip=]{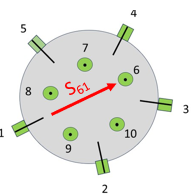}
\end{tabular}
\end{center}
 \caption{\small Antenna design (Ansys HFSS).  Right figure shows numbering of antennas and  scattering parameter $S_{61}$: measure of transmission from antenna 1 to antenna 6. }
\label{fig:HFSS1}
\end{figure}

\begin{figure}[tbp]
 \begin{center}
   \begin{tabular}{c}
     \includegraphics[trim = 0.0cm 0.0cm 0.0cm 0.0cm, scale=0.5, clip=]{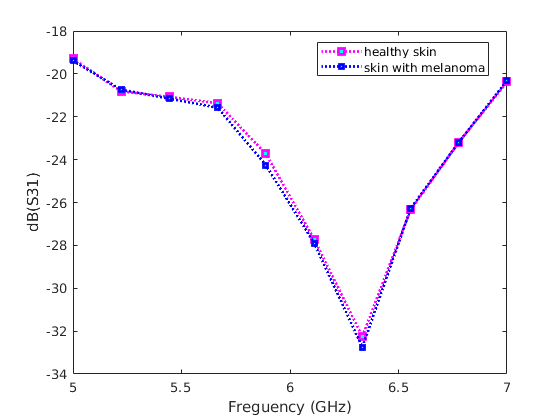}  \\
     a) \\
     \includegraphics[trim = 0.0cm 0.0cm 0.0cm 0.0cm, scale=0.5, clip=]{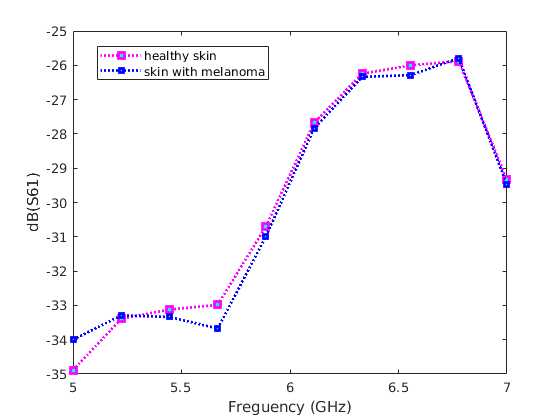} \\
     b) \\
     \includegraphics[trim = 0.0cm 0.0cm 0.0cm 0.0cm, scale=0.5, clip=]{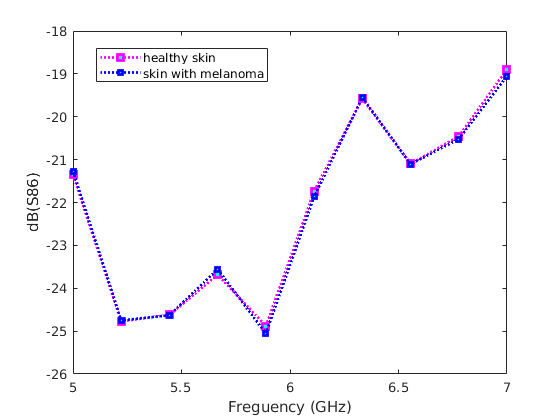} \\
     c)
\end{tabular}
\end{center}
 \caption{\small  Measured scattering parameters for antenna design shown
      on Figure \ref{fig:HFSS1}. }
\label{fig:HFSS2}
\end{figure}

\begin{figure}[tbp]
 \begin{center}
   \begin{tabular}{c}
     \includegraphics[trim = 0.0cm 0.0cm 0.0cm 0.0cm, scale=0.5, clip=]{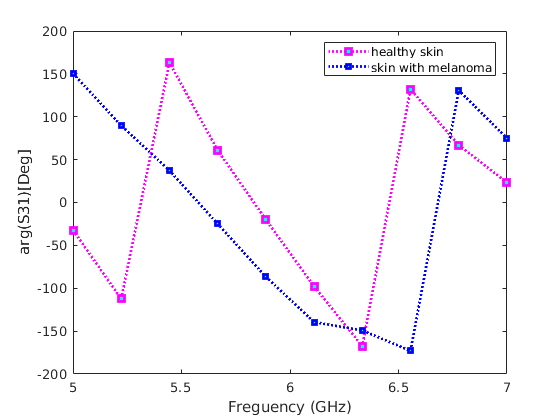} \\
       a) \\
       \includegraphics[trim = 0.0cm 0.0cm 0.0cm 0.0cm, scale=0.5, clip=]{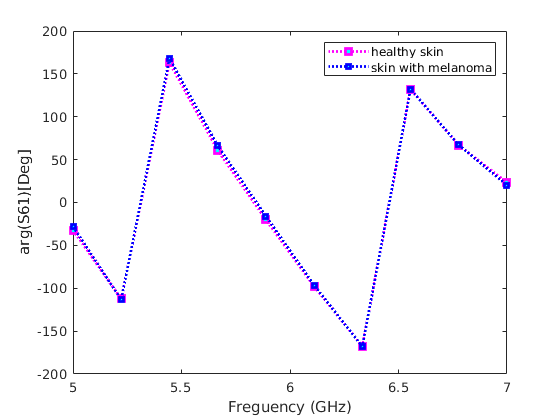}  \\
       b) \\
       \includegraphics[trim = 0.0cm 0.0cm 0.0cm 0.0cm, scale=0.5, clip=]{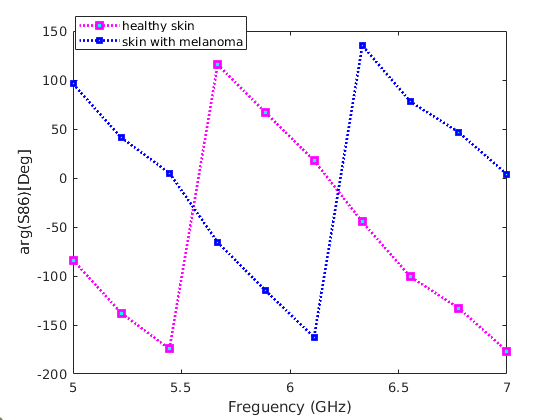} \\
       c)
\end{tabular}
\end{center}
 \caption{\small Measured scattering parameters for antenna design shown on Figure \ref{fig:HFSS1}. }
\label{fig:HFSS3}
\end{figure}

\begin{figure}[tbp]
 \begin{center}
   \begin{tabular}{cc}
     \includegraphics[trim = 0.0cm 0.0cm 0.0cm 0.0cm, scale=0.6, clip=]{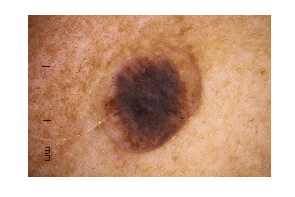} &
         \includegraphics[trim = 0.0cm 0.0cm 0.0cm 0.0cm, scale=0.5, clip=]{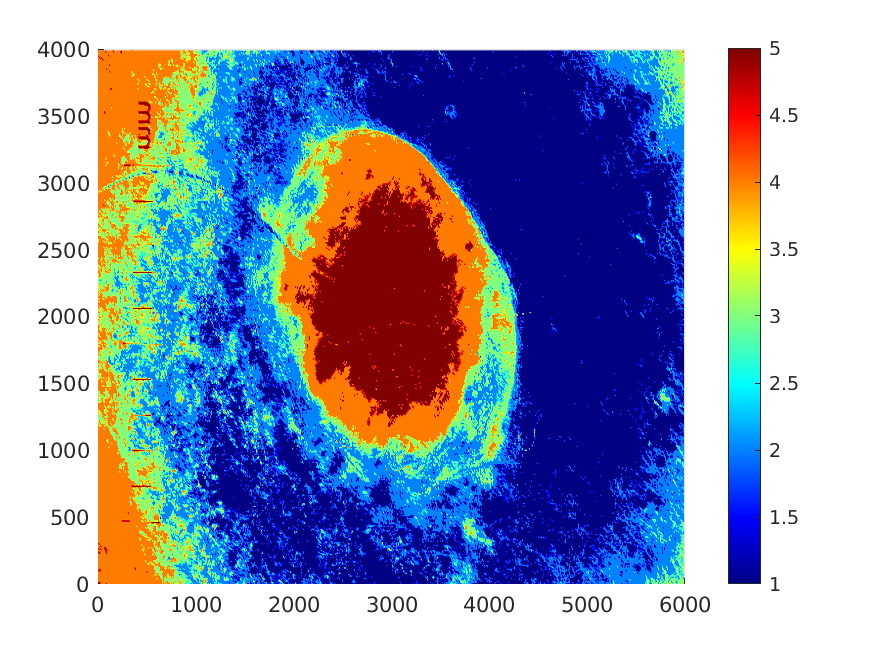} 
\end{tabular}
\end{center}
 \caption{\small
Left:   Image from ISIC dataset \cite{isic} taken for analysis. Right: segmentation of ISIC image into 5 types of material. }
\label{fig:ISIC2}
\end{figure}

\begin{figure}[tbp]
 \begin{center}
   \begin{tabular}{cc}
     \includegraphics[trim = 0.0cm 0.0cm 0.0cm 0.0cm, scale=0.5, clip=]{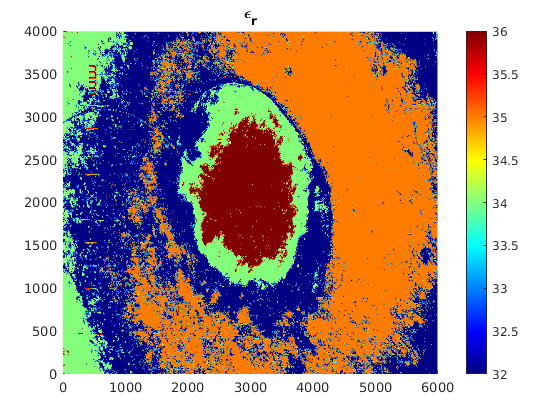} &
         \includegraphics[trim = 0.0cm 0.0cm 0.0cm 0.0cm, scale=0.5, clip=]{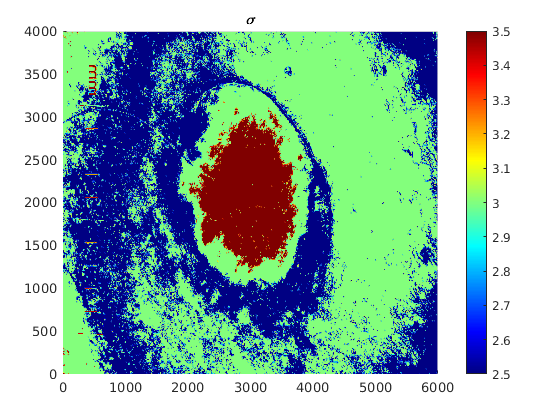} 
\end{tabular}
\end{center}
 \caption{\small Modelled  dielectric properties of skin  for ISIC image of Figure \ref{fig:ISIC2}. Left image: $\varepsilon_r$, right image: $\sigma$.}
\label{fig:ISIC3}
\end{figure}

\begin{figure}[tbp]
 \begin{center}
   \begin{tabular}{c}
     \includegraphics[trim = 4.0cm 0.0cm 0.0cm 0.0cm, angle = 0,scale=0.4, clip=]{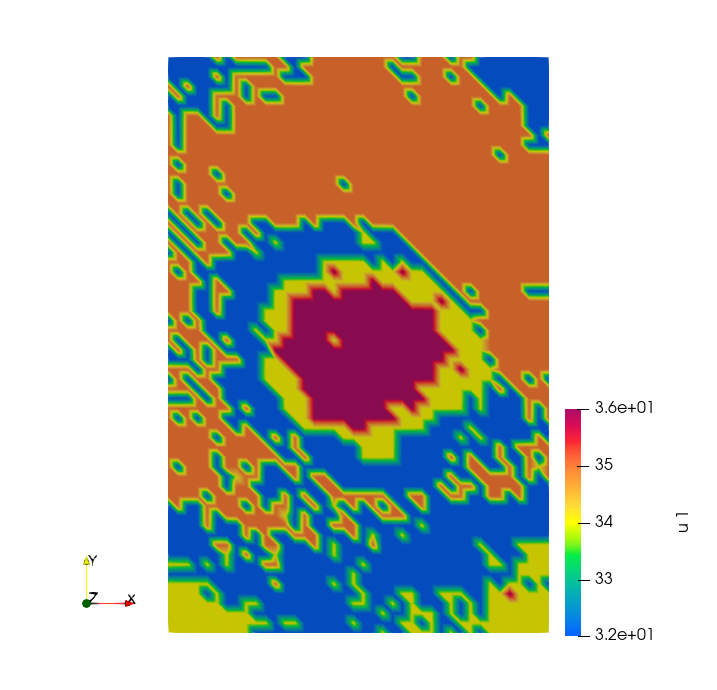} \\
     a) \\
       \includegraphics[trim = 0.0cm 0.0cm 0.0cm 0.0cm, angle = 0,scale=0.5, clip=]{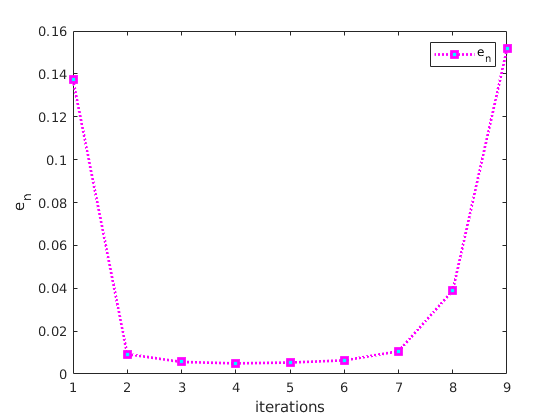}\\
       b)
\end{tabular}
\end{center}
 \caption{\small
 Performance of the Algorithm 1:  reconstruction of  $\varepsilon'$ with known  $\sigma$ of analyzed ISIC image 
   for realistic values of dielectric properties of skin: 
  a) reconstructed relative dielectric permittivity function $\varepsilon_r$; b) convergence of Algorithm 1.
 }
\label{fig:ISIC4}
\end{figure}

\section{Numerical Studies}\label{S4}
\label{sec:num}

\begin{table}[ht]
    \centering
    \begin{tabular}{|c|c|c|c|c|}
    \hline
                &  6 GHz & Test 1 &  Test 2 &  depth   \\
    Tissue type  & $\varepsilon_r$  & $\sigma$  (S/m) & $\sigma$  (S/m) & (mm) \\
    \hline
    Immersion medium       & 32 & 4 & 0.4 &  2   \\
    \hline
    epidermis              & 35 & 4  & 0.4   & 1  \\
    \hline
     dermis                & 40 &  9  & 0.9 & 3.5  \\
    \hline
    Fat                    & 9  &  1   & 0.1 &  5.5  \\ 
    \hline
    Tumor stage 1          & 45 &  5  & 0.5 & $< 1$  \\
    \hline
    Tumor stage 2          & 50 &   5  & 0.5 & $ > 1$  \\
    \hline
    Tumor stage 3          & 60 &  6   & 0.6 &  $> 1$  \\
    \hline
    \end{tabular}
     \caption{\textit{Tissue types and corresponding  realistic
      values of $\varepsilon_r$  and $\sigma$  (S/m) at 6Ghz for  skin with melanoma
      used in our numerical experiments.}}
    \label{tab:table1}
\end{table}

In this section we present results of the ongoing work on data
acquisition procedure using prototype of the device for MM detection
in 3D.  Additionally, we demonstrate performance of the Algorithm 1
using simulated data in 2D for the reconstruction of the function
$\varepsilon_r$ in \eqref{rotrot2} in the simplified case when we know
exact values of $\sigma$.

In both cases we are using realistic values of dielectric properties
of skin at 6 Ghz taken from \cite{mri}. Table \ref{tab:table1}  shows
the dielectric properties of healthy skin and melanoma at 6Ghz based on Figure 2 of \cite{mri}
which are used in our computational setup.
Thickness for different stages of melanoma (in mm) in this table is given accordingly  \cite{skindepth}.

\subsection{Data acquisition procedure}

 For data acquisition in 3D we
  set the computational domain $\Omega$   as
  \begin{equation*}
    \Omega = \left\{ x= (x_1,x_2, x_3) \in
           (0,  10)
    \times (0,  10)
    \times (0, 10)(mm)
    \right\},
 \end{equation*}
  and assign values of $\varepsilon_r$ and $\sigma$ accordingly to the Table
\ref{tab:table1}
  - see details on Figure
  \ref{fig:model3D}.

  We tested
 different values of $\sigma$  presented in the  Table \ref{tab:table1}.
 Figures
 \ref{fig:HFSS2},  \ref{fig:HFSS3}
  present results of measurements
  for $\sigma$
   corresponding to the Test 2 of the Table \ref{tab:table1}.
   Figure \ref{fig:HFSS2}-a) shows the scattering parameter S31 (in dB)
   when melanoma is present (blue) and when melanoma is absent
   (red). This is a measure of the transmission from horizontal
   antenna 1 to horizontal antenna 3, see Figure \ref{fig:HFSS1}.
  A difference of up to 0.5 dB is noticed between melanoma and
   ”healthy” skin.

Figure \ref{fig:HFSS2}-b) shows the scattering parameter S61 (in dB) when melanoma is present (blue) and when melanoma is absent (red). This is a measure of the transmission from horizontal antenna 1 to vertical antenna 6, see Figure \ref{fig:HFSS1}.  A difference of up to 0.9 dB is noticed between melanoma and ”healthy” skin.

Figure \ref{fig:HFSS2} -c) shows the scattering parameter S86 (in dB) when melanoma is present (blue) and when melanoma is absent (red). This is a measure of the transmission from vertical antenna 6 to vertical antenna 8, see Figure \ref{fig:HFSS1}.  A small difference of up to 0.1 dB is noticed between melanoma and ”healthy” skin.

Figure \ref{fig:HFSS3}-a)  shows the phase angle (arg(S31)) of scattering parameter S31 (in degrees) when melanoma is present (blue) and when melanoma is absent (red). This is a measure of the phase of transmission from horizontal antenna 1 to horizontal antenna 3, see Figure \ref{fig:HFSS1}.
It is noticed a significant phase distortion when melanoma is present.

Figure \ref{fig:HFSS3}-b) shows the phase angle of scattering parameter S61 (in degrees) when melanoma is present (blue) and when melanoma is absent (red). This is a measure of the phase of transmission from horizontal antenna 1 to vertical antenna 6.
It is noticed a negligible phase distortion when melanoma is present, in contrast to the significant magnitude difference seen in Figure  \ref{fig:HFSS2}-b) .

Figure  \ref{fig:HFSS3}-c) shows the phase angle of scattering parameter S86 (in degrees) when melanoma is present (blue) and when melanoma is absent (red). This is a measure of the phase of transmission from vertical antenna 6 to vertical antenna 8.
It is noticed a significant phase distortion when melanoma is present similar to Figure \ref{fig:HFSS3}-a).

Note: arg(S) is analogous to arg(z). The argument is the angle between the vector z and the real x-axis.

  \subsection{Performance of Algorithm 1 in 2D}

  This section demonstrates study of performance of the Algorithm 1 in 2D
   case 
  using simulated data of the electric field. For generation of simulated data we used image
  from ISIC dataset \cite{isic}  presented on the left image of Figure
  \ref{fig:ISIC2}.  The skin image was segmented into 5 types of
  materials depending on the pixels value - see right image of Figure
  \ref{fig:ISIC2}. Next, corresponding to every type of material we
  assigned values of $\varepsilon_r$ and $\sigma$ as it is shown on
  Figure \ref{fig:ISIC3}. Since  analyzed  skin image of Figure
  \ref{fig:ISIC2} is benign
  its dielectric properties align with those typically associated with benign tissue
  -  compare  values of images in Figures \ref{fig:ISIC3}  with  dielectric properties
  of  skin presented in Table \ref{tab:table1} ( see $\sigma$ for Test 1).

  We tested Algorithm 1 in the simplified case when $\sigma=0$.
  Thus,
  the equation \eqref{2.7n} will contain only values of
  $\varepsilon_r$  and transforms to the real valued equation.
 To speed-up computations we
  sampled original computational mesh of the left Figure
 \ref{fig:ISIC2} which has size $4000 \times 6000$, to the 
   mesh of the size
  $40 \times 60$.
  
  For  generation of data $\tilde{E}$  in equation \eqref{2.7n}  we used domain
  decomposition method of \cite{BL1} in 2D, see details for generation of data
  in \cite{BGG}.  We added normally distributed  Gaussian  noise  $\delta= 5\%$ 
  to generated data  $\tilde{E}$  and then smoothed obtained data - see details of this procedure
  in \cite{BL2,BGG}.

Figure  \ref{fig:ISIC4}  shows computational results
of the reconstruction of the function $\varepsilon_r$
 using  Algorithm 1. We  observe that the algorithm converged
 at iteration $n=6$ and the computed  values of the related dielectric permittivity  correspond very good to the
 exact values of the
  original function $\varepsilon_r$  presented on the left figure of
 Figure \ref{fig:ISIC3}.

\section{Conclusions}\label{S5}

\label{sec:concl}

In this work we develop  variational  iterative algorithm for reconstruction of the
space dependent complex dielectric permittivity $\varepsilon'(x)$
using   back-scattered measurements
of the electric field in frequency domain.
  Our numerical test show efficiency and robustness of the proposed 
  algorithm in the case of the known  conductivity function.

  Additionally, we present data acquisition procedure using prototype
  of the device for MM detection
  modelled in Ansys HFSS.
  Our current work is concentrated on testing of  data  measured in HFSS Ansys 
  using proposed algorithm in order to detect MM on different stages.


\end{document}